# In search of truth: Evaluating concordance of AI-based anatomy segmentation models

Lena Giebeler[1,2], Deepa Krishnaswamy[2], David Clunie[3], Jakob Wasserthal[4], Lalith Kumar Shiyam Sundar[5], Andres Diaz-Pinto[6], Klaus H. Maier-Hein[7], Murong Xu[8], Bjoern Menze[8], Steve Pieper[9], Ron Kikinis[2], Andrey Fedorov[2]*

** corresponding author, fedorov@bwh.harvard.edu*

*[1]RWTH Aachen University, Aachen, Germany*
*[2]Brigham and Women's Hospital, Boston, MA, USA*
*[3]PixelMed Publishing, Bangor, PA, USA*
*[4]University Hospital of Basel, Switzerland*
*[5]Ludwig Maximilian University Munich, Germany*
*[6]NVIDIA, USA*
*[7]German Cancer Research Center (DKFZ), Germany*
*[8]University of Zurich, Switzerland*
*[9]Isomics Inc, Cambridge, MA, USA*

## Abstract

**Purpose** AI-based methods for anatomy segmentation can help automate characterization of large imaging datasets. The growing number of similar in functionality models raises the challenge of evaluating them on datasets that do not contain ground truth annotations. We introduce a practical framework to assist in this task. **Approach** We harmonize the segmentation results into a standard, interoperable representation, which enables consistent, terminology-based labeling of the structures. We extend 3D Slicer to streamline loading and comparison of these harmonized segmentations, and demonstrate how standard representation simplifies review of the results using interactive summary plots and browser-based visualization using OHIF Viewer. To demonstrate the utility of the approach we apply it to evaluating segmentation of 31 anatomical structures (lungs, vertebrae, ribs, and heart) by six open-source models - TotalSegmentator 1.5 and 2.6, Auto3DSeg, MOOSE, MultiTalent, and CADS - for a sample of Computed Tomography (CT) scans from the publicly available National Lung Screening Trial (NLST) dataset. **Results** We demonstrate the utility of the framework in enabling automating loading, structure-wise inspection and comparison across models. Preliminary results ascertain practical utility of the approach in allowing quick detection and review of problematic results. The comparison shows excellent agreement segmenting some (e.g., lung) but not all structures (e.g., some models produce invalid vertebrae or rib segmentations). **Conclusions** The resources developed are linked from https://imagingdatacommons.github.io/segmentation-comparison/ including segmentation harmonization scripts, summary plots, and visualization tools. This work assists in model evaluation in absence of ground truth, ultimately enabling informed model selection.

## 1. Introduction

Growing public availability of large imaging datasets offers distinct opportunities for secondary analysis and hypothesis exploration. Public imaging datasets that are readily available and actively utilized in the community for secondary research studies often include data collected from thousands of subjects. Prominent examples of the completed and ongoing data collection initiatives include The Cancer Genome Atlas (TCGA)[1] (images for n>11,000 subjects), National Lung Screening Trial (n>26,000)[2], Childhood Cancer Data Initiative (n>4800)[3], to name just a few. Cloud-based tools to search, download and analyze these large datasets, such as those available via National Cancer Institute Imaging Data Commons (IDC)[4] amplify opportunities for discovery. As a prerequisite, secondary analyses of such datasets often require extraction of quantitative measurements characterizing anatomy and pathology. Until recently, studies involving quantitative analysis of images for thousands of cases were not feasible due to enormous effort needed for expert annotation of the images, and lack of robust tools for automated image analysis. Emerging robust AI-based methods for anatomy segmentation[5] can help automate characterization of large imaging datasets by providing the anatomical context needed for analyses that cannot be achieved from images and their metadata alone. Such secondary analyses can include correlations of image-derived features with clinical outcomes or genomic/proteomic markers, construction of cohorts based on anatomical characteristics, and identification of similar cases across populations[6]. Segmentations also enable direct measurement of organ volumes, shapes, and spatial relationships, providing valuable insights for disease characterization, treatment planning, and population-level studies[5,7].

AI-based segmentation approaches offer the ability to generate large numbers of annotations with considerably less expert effort[5]. Early AI-based segmentation methods were typically developed for specific tasks, focusing on individual organs or structures[5]. More recently, a growing number of whole-body segmentation models have been developed to automatically segment multiple anatomical structures at once, with *TotalSegmentator* being one of the first such models[8]. Availability of such robust segmentation tools opens new opportunities for processing and quantitative analysis of large imaging datasets, which are not practical to annotate manually. Imaging Data Commons (IDC)[4] is a cloud-based environment hosting a growing number of publicly available cancer imaging collections and tools to support the use of the data. While IDC provides access to versatile imaging data, most of it is not accompanied by segmentations or quantitative features, limiting its use for secondary analyses. Adding AI-based segmentations could streamline reuse of these datasets[9–13]. Given the non-trivial effort and computational cost needed to apply AI segmentation to a large cohort, storage implications and potential for the confusion by the users, it is important to investigate selection of a preferred segmentation tool among alternatives.

Our study builds on the earlier work by Thiriveedhi et al, where *TotalSegmentator* 1.5 was applied to produce segmentations for NLST[5,14]. Since then, several newer whole-body segmentation models have been released, raising the question of whether any of these models may outperform *TotalSegmentator*. We develop a framework to enable systematic comparison of segmentations produced by different models, and apply it to evaluate six segmentation models on a sample of images from the National Lung Screening Trial (NLST) collection[15], one of the largest in IDC. We demonstrate the utility of the framework by comparing results generated by *TotalSegmentator* 1.5[8,16], *TotalSegmentator* 2.6[8,16], *Auto3DSeg*[17], *MOOSE*[18,19], *MultiTalent*[20,21], and *CADS*[22,23] models. We limit our evaluation to the open source segmentation models that do not impose any constraints on reuse and distribution of the segmentation results (including commercial re-use), since our goal is to share the resulting stored segmentations via IDC with minimal restrictions. Our objective is to identify the most robust model, which could then be used to generate segmentations for the entire NLST cohort and make them publicly available through IDC. The comparison process and accompanying tools we establish are designed to be transferable, allowing it to be applied to any other 3D imaging dataset, both within and outside of IDC.

Selecting the most suitable model for the NLST dataset is not straightforward. The NLST dataset lacks ground truth segmentations, making it difficult to objectively evaluate model performance. Expert review of hundreds of structures across thousands of scans is not practical. The outputs of segmentation tools are not standardized: the same anatomical structures may be assigned different labels or their boundaries defined differently, complicating direct comparison. There is also a paucity of tools that support efficient visual comparison of segmentations across models.

Earlier studies that compared some of the models we selected relied on the availability of reference annotations and report performance based on comparisons with manual ground truth. Among the models included in our study, *TotalSegmentator* was the first model to segment >100 structures in CT,, and therefore often serves as a benchmark and surrogate for ground truth in related work. Auriac et al.[24] compared *TotalSegmentator* v2.0.5 and *MOOSE* on PET/CT data, focusing exclusively on organ-level segmentations. They reported moderately close agreement between the two models for large organs, with volume differences below 10% in at least 80% of patients. Amini et al.[25] reported slightly higher lobar segmentation accuracy for *TotalSegmentator* in chest CT, in a study that also included *MOOSE* and *LungMask*[26] models evaluated against expert segmentations. Similarly, the *CADS* developers demonstrated that their model achieved comparable performance to *TotalSegmentator* on body CT data, while providing more consistent bone and soft-tissue boundary delineations[22,23].

When ground truth is absent, prior work has explored several strategies for model evaluation. These include reverse classification accuracy (RCA)[27], where a predicted segmentation is used as a pseudo-label to infer its own quality, as well as regression or transformer-based quality prediction models that estimate segmentation accuracy directly from image–mask pairs[28,29]. However, these approaches require retraining of the model being evaluated. Alternatively, consensus-based and benchmarking workflows have been proposed to compare multiple segmentation models on the same data without relying on ground truth. Examples include majority voting and STAPLE fusion[30], which generate ensemble consensus segmentations, and the weighted ensemble agreement framework by Sims et al.[31], which evaluates models relative to a consensus derived from all predictions.

We address the challenges of model evaluation by (1) harmonizing the storage format of the encoded segmentation results, (2) developing open-source tools in 3D Slicer for streamlined visualization and comparison, (3) quantifying inter-model agreement using consensus-based metrics, and (4) delivering a browser-based interface to examine segmentation results for the individual images. By systematically comparing functionally similar models on a sample of CT images from the NLST collection, our study demonstrates early results in identifying differences across the evaluated models, and contributes practical guidance for informed model selection in absence of expert annotations.

# 2. Methods

To systematically evaluate and compare multiple AI-based anatomical segmentation models in the absence of ground truth data, we first harmonized the segmentation outputs from all six open-source models on our evaluation dataset into DICOM SEG format[32]. This made it straightforward to identify segments corresponding to the same structures, visualize those using consistent color, and also enabled interoperability with the existing DICOM archives and viewers. We then performed both quantitative and qualitative analyses.

We started with quantitative analysis, assessing inter-model agreement using consensus segmentation (portion of the segmentation of a given structure where all models agreed). We then evaluated pair-wise per-structure (volumetric) agreement for each of the models with the consensus segmentation using the Dice Similarity Coefficient (DSC)[33] and the volume. Since the analysis produced a large amount of data, with one DSC and one volume value per model and structure, we visualized the results in interactive plots. Each data point in the plots links directly to the corresponding CT scan and its associated segmentations in the OHIF Viewer[34].

We then selected all structures with less than 90% agreement (in either DSC or volume) for further visual inspection. These cases formed the basis of our qualitative analysis, with priority given to those showing the largest deviations in the interactive plots. OHIF was used to pre-select notable examples, which were then examined in detail using 3D Slicer[35]. To support this, we developed *CrossSegmentationExplorer* - an interactive 3D Slicer module that allows side-by-side comparison of segmentations across models[36]. Finally, selected findings from the visual analysis were reviewed by an expert with the training in radiology and specialized expertise in volumetric anatomy segmentation (R.K.).

The following sections provide a detailed overview of each step in the analysis.

## 2.1. Evaluation dataset

Our initial evaluation dataset consists of two components.

We used a sample of the CT images available publicly as part of the NLST[2,37] collection in The Cancer Imaging Archive, one of the largest publicly available imaging datasets for cancer screening, consisting of chest CT scans. For our analysis, we selected 18 DICOM CT series from four patients within the NLST collection, covering a total of nine DICOM studies (a study corresponding to a single image acquisition session). The specific sample analyzed was selected by the lead developer of the MOOSE model from the NLST collection for the purpose of demonstrating the operation of the MOOSE model. This sample initiated the investigations of how that model compared with the *TotalSegmentator* 1.5 results that were already in IDC at the time. As we were discovering new models, we applied those to the same sample. An overview of the selected case composition and detailed acquisition parameters are provided in Table 1 in the Appendix.

We also utilized volumetric segmentations of the CT scans produced earlier using *TotalSegmentator* v1.5[38] and shared as part of the TotalSegmentator-CT-Segmentations collection[10,39] in IDC.

Both images and baseline *TotalSegmentator* segmentations were obtained from IDC, which provides public access to the NLST imaging dataset and a subset of the associated clinical metadata. The corresponding metadata were obtained using the IDC Google BigQuery interface.

## 2.2. Segmentation models

The segmentation models used in this study differ in terms of their architecture, training data, and anatomical coverage. Most are based on nnU-Net or MONAI pipelines and have been trained on diverse public and institutional datasets[8,17,19,21].

Despite these differences, they have in common that all models used in this study produce their segmentation outputs in the NIfTI format. Below, we briefly summarize the key characteristics of each model. Importantly, we considered only those models of *TotalSegmentator* that did not require a license, and did not restrict commercial usage (all of the other models do not have any components that have such restrictions).

- ***TotalSegmentator* 1.5** (https://github.com/wasserth/TotalSegmentator) is a multi-organ segmentation model for CT images released in September 2023. It is based on the nnU-Net architecture and segments 104 anatomical structures across the entire body. The model was trained on 1,204 clinical CT scans from the University Hospital Basel, collected in 2012, 2016, and 2020[8,16].
- ***TotalSegmentator* 2.6** (https://github.com/wasserth/TotalSegmentator) is an extended version released in January 2025 that supports both CT and MR images. The updated total task for CT includes 117 anatomical structures. In addition, 20 open-source tasks have been added for the targeted segmentation of specific anatomical regions and pathologies (e.g., head and neck, liver segments, vessels, implants). Across all tasks, the model can segment up to 238 CT and 85 MR structures[16].
- ***Auto3DSeg*** (https://github.com/Project-MONAI/tutorials/tree/main/auto3dseg) is an automated segmentation framework built on MONAI[40] that constructs full 3D segmentation pipelines with minimal user input. For our analysis, we used an *Auto3DSeg* model that was trained on the publicly available

*TotalSegmentator* dataset. This resulted in the segmentation of 117 anatomical structures in CT images[17].

- ***MultiTalent*** (https://github.com/MIC-DKFZ/MultiTalent), released in May 2023[20,21], is best understood as a large-scale multi-dataset training strategy rather than a single public model. In this work, a MultiTalent model pretrained on more than twenty thousand volumes from 65 public datasets across CT, MRI and PET was used. It was then fine-tuned on the TotalSegmentator dataset for 1000 epochs to align with the target anatomical structures. Full details on pretraining datasets and training can be found in[41].
- ***MOOSE* 3.0** (https://github.com/ENHANCE-PET/MOOSE), released in November 2024, is a large-scale multi-organ segmentation model trained on 1,683 low-dose whole-body and total-body CT ENHANCE.PET dataset[42]. It segments 130 anatomical structures, using specialized clinical submodels focused on abdominal organs, muscles, bones, cardiac subregions, vessels, adipose tissue, and skeletal muscle around the third lumbar vertebra. The model builds on a 3D nnU-Net architecture and incorporates a data-centric training strategy to optimize performance across heterogeneous datasets[18,19]. The training dataset includes images from individuals without overt disease and patients with a range of malignant and inflammatory pathologies, including arthritis, lymphoma, and melanoma, as well as cancers of the lung, head-neck, and genito-urinary tract.
- ***CADS*** (https://github.com/murong-xu/CADS/tree/main), released in July 2025, is an open-source multi-organ segmentation model for CT images trained on the *CADS*-dataset, which contains 22,022 CT scans from over 100 centers in 16 countries. Importantly, this model was trained using 7172 CT scans from the NLST collection. It segments 167 anatomical structures from head to knees using a nnU-Net–based architecture with region-specific submodels. The model was validated on 18 public datasets and a large real-world oncology cohort[22,23].

All of the models except *MOOSE* used the *TotalSegmentator* CT dataset for training[38] (with the labels refined before training the *CADS* model). Further, both *CADS* and *MultiTalent* utilized the *AMOS*[43], *KiTS19*[44] and *BTCV-Abdomen*[45] datasets.

## *2.3. Segmentation results harmonization*

We converted all segmentation outputs to the DICOM SEG format, which supports standard anatomic coded terms for describing segmentation results, provides a standard means for encoding assigned colors recommended for visualizing the segments, and encodes references to the segmented images[32].

The semantics of a segment are defined using three DICOM SEG coded attributes: the category, which indicates the general context (e.g., “anatomical structure”); the type, which specifies the anatomical target (e.g., “liver” or “lung”); and an optional type modifier, which adds detail such as laterality (“left” or “right”), if not pre-coordinated, or anatomical variants. For segments that do not correspond to a specific anatomic structure, such as body fat, the category is assigned to be “tissue”. Segmentations of the entities that are not anatomic organs (e.g., cyst, tumor, implant) can include anatomic region code assignment to communicate spatial location of the entity. Each of these coded attributes is represented as a triplet consisting of the coding scheme (the coding system), the code value (the coded identifier specified by the coding system), and the code meaning (human-readable text label in the local language, used as the displayed name). We used the SNOMED-CT[46,47] coding system - an internationally recognized system for medical terminology, which is the terminology preferred by DICOM for describing segmentation result properties[48].

To create the mapping from model-specific labels to SNOMED-CT terms, we started with the mapping of labels to SNOMED-CT codes and colors that was done for *TotalSegmentator* earlier[10,49]. To complete the mapping, we defined DICOM visualization colors for the 61 structures that did not have color assignment. To select the initial colors, we used a Perplexity.AI large language model (LLM) assistant to assign visually distinct colors to the individual structures, which were then reviewed manually. The resulting mapping file was then used as the initial *TotalSegmentator* harmonized label map.

For the remaining segmentation models (*MOOSE*, *CADS*, *MultiTalent*, *Auto3DSeg*, and *TotalSegmentator* 2.6.), we applied an iterative procedure of harmonizing all of the labels across those models, and updating the harmonized label map:

1. **Label matching:** For *MOOSE* and *CADS*, label names were manually matched to the corresponding entries in the *TotalSegmentator* harmonized label map when possible. *MultiTalent*, *Auto3DSeg*, and *TotalSegmentator* 2.6, on the other hand, reused or extended the same label names as *TotalSegmentator* 1.5. Therefore, whenever a structure was already present in the *TotalSegmentator* harmonized label map, it was assigned the same label name. Based on these matches, the corresponding SNOMED-CT codes and DICOM colors from the *TotalSegmentator* 1.5 mapping were reused.
2. **Cross-check with other completed mappings:** For labels not found in the *TotalSegmentator* harmonized label map, we attempted to match them against already completed mappings from already processed models.
3. **Manual SNOMED-CT code assignment:** If no match was found for a structure, the SNOMED-CT codes were added manually by reviewing the anatomical meaning of the label in the associated publications and mapping those to the code using the online SNOMED-CT browser (https://browser.ihtsdotools.org/).
4. **SNOMED-CT code comparison:** After processing all labels, we compared the newly assigned SNOMED CT codes with those in other finalized mappings to identify matching structures that might not have been apparent through label names alone. If a match was found, we reused the color already assigned for the corresponding structure in the other mapping.
5. **Color assignment:** Finally, any missing colors were added.

Finally, we converted the model-specific segmentations into DICOM. For this, we used the NIfTI model outputs and the corresponding harmonized metadata as an input into the *dcmqi*[50,51] pipeline to generate six DICOM SEG files for each CT scan series in our evaluation dataset, one for each segmentation model. The conversion scripts are available in https://github.com/ImagingDataCommons/segmentation-comparison.

## *2.4. Segmented structure selection*

To enable a meaningful comparison between models and reduce the number of structures considered, we applied a series of three exclusion criteria to refine the set of structures included in the study. First, we excluded structures that were only segmented by a single model, since no inter-model comparison was possible for these. Second, we reviewed segmentation results visually and removed structures with incomplete coverage in the chest CT scans, as this often resulted in unreliable or anatomically implausible segmentations. Not surprisingly, those lung screening scans often did not have complete coverage for abdominal organs, such as liver, or neck area. Segmentation models are typically trained on complete organ volumes, so the absence of important spatial context limits their ability to perform reliably. Finally, we excluded structures segmented by only two or three models. Since no ground truth segmentations were available, our evaluation relied on inter-model agreement. While it is possible that all models could be wrong, the probability of a segmentation being correct increases when multiple independent models agree on the same structure. Therefore, structures supported by only a small subset of models were excluded from further comparison.

## *2.5. Quantitative evaluation of agreement/concordance*

In the quantitative analysis, we evaluated how consistently each anatomical structure was segmented across the six models. As a first step, we computed the consensus for each structure. The consensus represents the set of voxels that were segmented by all models that include this structure. We then compared each model's segmentation to the consensus using two metrics to quantify agreement: the DSC and the volume. The DSC is a widely used metric[52] that quantifies the spatial overlap between two binary segmentation masks. In our analysis, we used the DSC to measure the similarity between the segmentation of each model and the corresponding consensus for each structure.

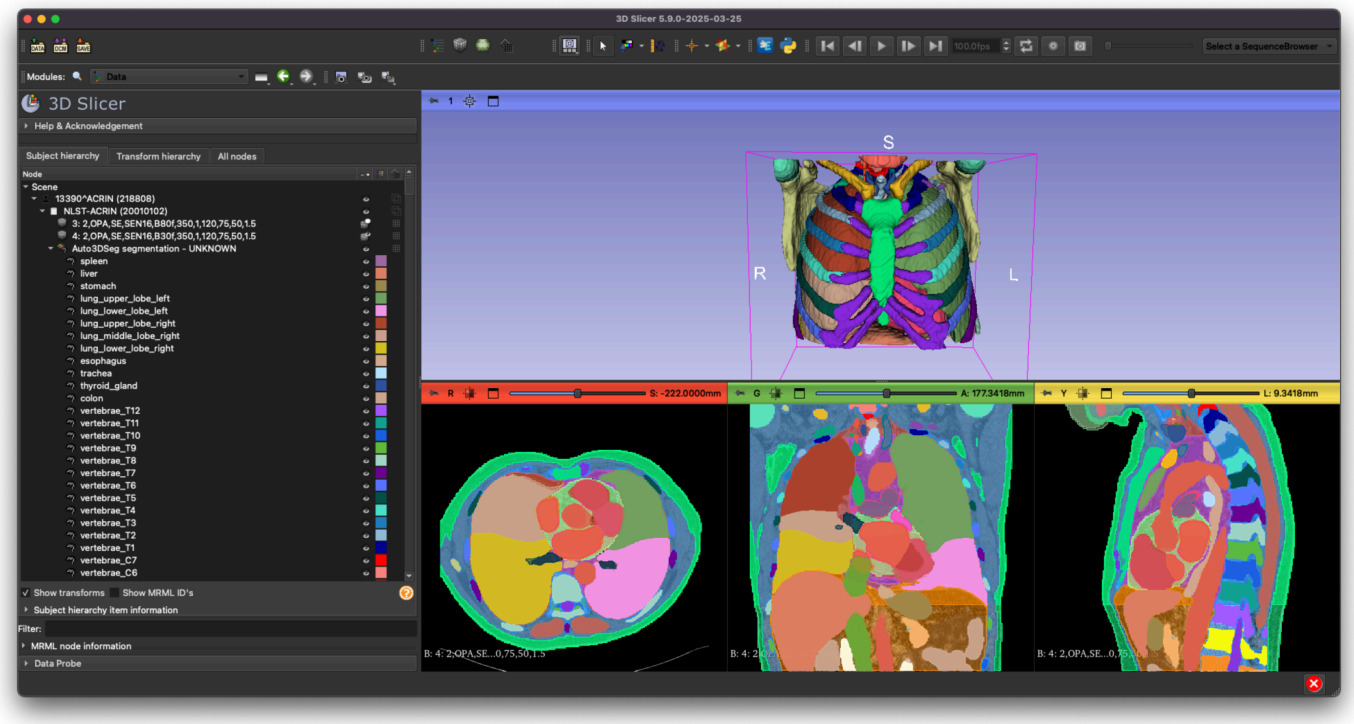

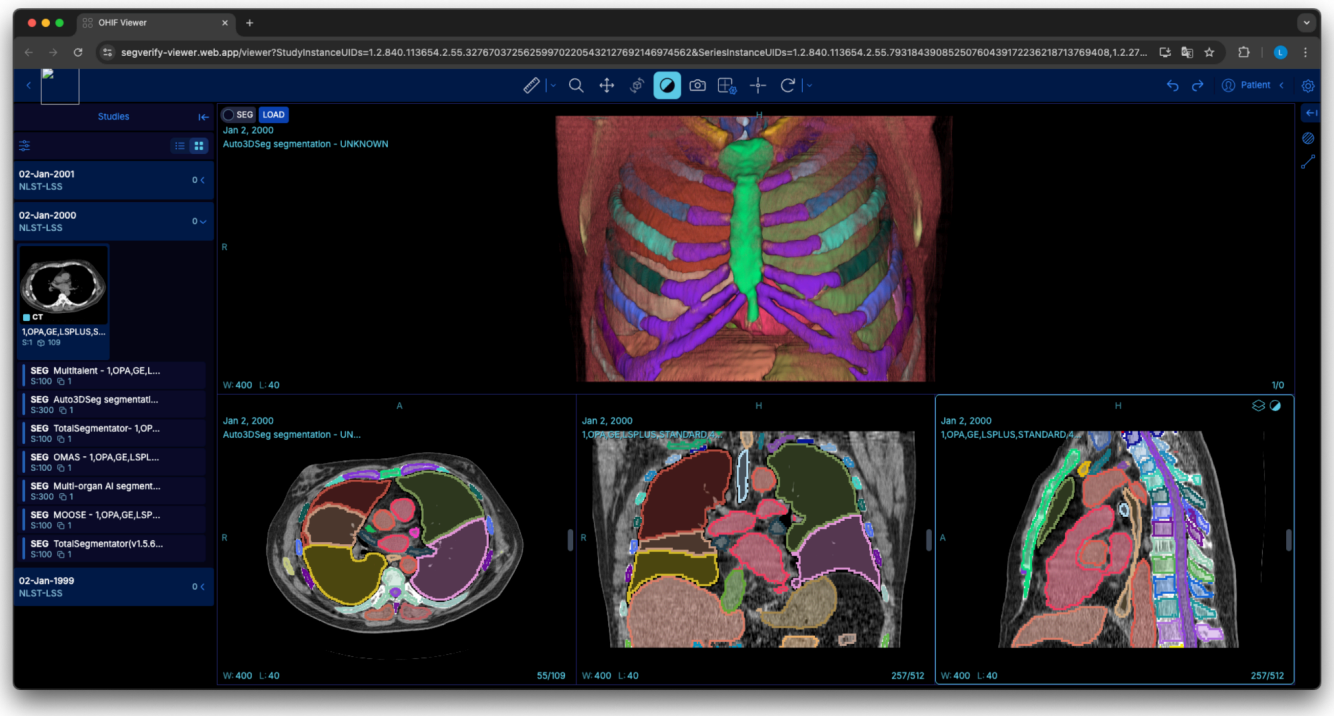

***Figure 1:*** *Example layouts of 3D Slicer (left) and OHIF (right) after loading a CT image and the corresponding DICOM-SEG files. In both cases, the layout shown consists of a 3D view on top and three 2D views (axial, coronal, sagittal) below. In 3D Slicer, this layout is the default after loading, and all loaded segmentations are automatically displayed in the 2D views. In OHIF, the layout must first be selected, and only the CT image is shown by default until a segmentation is explicitly loaded. In both tools, an Auto3DSeg segmentation is rendered in the 3D view as an example. In OHIF, the 3D view by default shows the CT volume rendering. Once a segmentation is selected, it is also displayed. The CT volume rendering can be hidden to view only the segmentation rendering.*

For volume-based analysis, we computed the volume for each segmented structure, as well as for the corresponding consensus segmentation. We then calculated the volume ratio between the consensus segmentation and each of the individual model segmentations. This comparison allowed us to assess how closely the models agreed in terms of volume estimation. In addition to analyzing per-structure agreement, we also analyzed whether certain models consistently produced larger or smaller segmentation volumes for specific structures across the dataset.

## *2.6. Summary visualization*

Applying the two metrics (consensus DSC and volume difference) across multiple CT scans and segmentation models quickly leads to a large number of data points. With 18 values per metric for each of the segments, judicious visualization of those is critical. To efficiently identify structures with strong disagreement among models within this large set of values, we developed customized interactive scatterplots, one for the consensus DSC and one for the volume-based analysis. These plots help to quickly identify outliers in model agreement for each structure. In both cases, we initially started with a standard non-interactive scatterplot and iteratively refined it until we developed a plot in which the outliers were clearly visible. The final plots are introduced in more detail in the Results section. By providing interactive plots it is possible to hover over individual points to access case-specific identifiers and to click on them to directly open the corresponding CT scan with all six segmentations in the OHIF Viewer (introduced in the next section).

### 2.7. Single-subject visualization

For the qualitative analysis, we performed visual comparisons using two open-source medical image visualization tools: *3D Slicer*[35,53,54] version 5.9.0 and the *OHIF Viewer*[34] version 3.11.0, which are both highly powerful tools that provide complementary capabilities for the task of comparing segmentation results. *OHIF Viewer*[34] (https://ohif.org) (Figure 1 right) is a web-based JavaScript thin-client image viewer that runs directly in the browser without the local installation of any software. It supports visualization of DICOM images and DICOM segmentations, making it well-suited for cloud-based workflows. One advantage of OHIF is that it enables direct access to public datasets hosted on IDC using the DICOMweb[55] API without requiring the DICOM data to first be locally downloaded. Users can also integrate visualization of their own datasets by uploading them to a Google Healthcare DICOM store and co-visualize them alongside public data (i.e., a single visualization can pull from multiple DICOMweb endpoints). As a result, once DICOM stores are populated with the images and segmentations, those can easily be shared using a structured URL and visualized, with the only prerequisite being a web browser. These features make OHIF Viewer a perfect tool for integrating with the interactive plots to enable quick single-subject visualization. 3D visualization, layouts and extensibility of OHIF, however, are rather limited, compared to those of 3D Slicer.

*3D Slicer*[35] (https://slicer.org) (Figure 1 left) is a desktop application for medical image analysis that supports clinical (DICOM) and research (NRRD and NIfTI) formats. It offers advanced tools for segmentation, multiplanar reconstruction, and 3D rendering. Segmentations (where each segmentation consists of one or more segments, or labels, in the typical AI segmentation model nomenclature) can be explored interactively across synchronized axial, sagittal, and coronal views, along with a linked 3D view. 3D Slicer offers unparalleled (among open source medical imaging software tools) flexibility in visualization, but requires local installation of the thick client, and (in a typical usage mode) local availability of the data. To support efficient review of segmentations generated using different models, we decided to develop a new *3D Slicer* extension that streamlines the process of loading and visualizing segmentations.

## 3. Results

### 3.1. Harmonized dataset

A total of 326 distinct anatomical structures can be segmented across all six models. Of these structures, 130 are segmented by more than one model: 91 structures are segmented by all six models; seven by five models; five by four models; 21 by three models; and six by two models. The remaining 196 structures are each segmented by only a single model. Specifically, *TotalSegmentator* 2.6 segments 116 structures exclusively; *CADS*, 53; and *MOOSE*, 27. The 91 structures segmented by all six models primarily cover central anatomical regions such as major organs (e.g., liver, lungs, kidneys), large blood vessels, the heart, and core components of the musculoskeletal system. Structures segmented by five to two models include smaller vascular branches, specific brain regions, and particular segments of the spine.

*TotalSegmentator* 2.6 is the only model that segments a wide range of additional anatomical structures, covering regions in the head and neck, musculature, thorax and abdomen, as well as skin and full-body compartments. *CADS* uniquely offers several head and neck structures, ENT-related regions, and selected soft tissue and vascular components, such as the salivary glands, pharyngeal spaces, and nasal cavities. *MOOSE* is the only model to segment paired bones and extremities, including hands, feet, and specific vertebral segments.

All of the segmentations were converted into DICOM SEG representation utilizing the harmonized mapping, resulting in 108 DICOM SEG files, one per model per segmented CT series. The resulting DICOM SEG files were imported into a Google Healthcare DICOM store[56], making them available through the DICOMweb interface for visualization using OHIF Viewer and integration with the interactive plots. Further, since Google Healthcare DICOM stores do not support public access, to enable public DICOMweb access to the data, we used Google CloudRun to deploy a simple proxy that provides unauthenticated access[57]. We then relied on the standard capabilities of OHIF Viewer to implement visualization of the CT images fetched from the IDC-maintained DICOMweb endpoint[58] and combine those with the SEG instances available via the proxy endpoint we established.

The harmonized dataset is publicly available[59], and includes the segmentations produced by the models, and the consensus segmentation. Along with the segmentations we provide the Excel spreadsheets that define mapping of the model-specific labels to SNOMED-CT codes and color assignment, as well as the detailed overview of all anatomical structures that can be segmented by the models, and the list of models that segment each structure.

### *3.2. Analyzed structures*

A total of 135 anatomical structures were segmented across all six models in our evaluation dataset. We excluded 29 structures because they were only segmented by a single model. 69 structures were removed due to their incomplete coverage in the chest CT scans. All abdominal organs fell into this category. Given the central role of the lungs in chest CT scans, an exception was made for the lower lung lobes, which were slightly cut off in one study. Lastly, we excluded 12 structures segmented by only two or three models. These structures included vascular components and smaller cardiac structures.

The final analysis was performed on 24 anatomical structures, which are shown in Figure 2: the thoracic vertebrae T2–T10, ribs 3–6, the sternum, both lungs (segmented as individual lobes), and the heart.

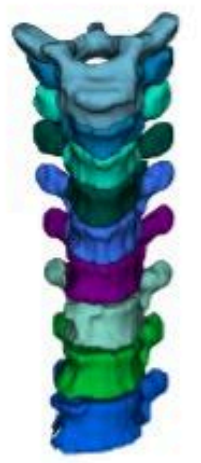
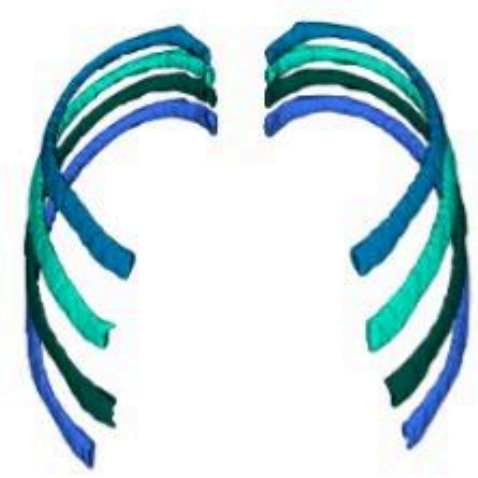
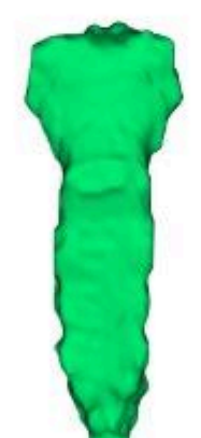
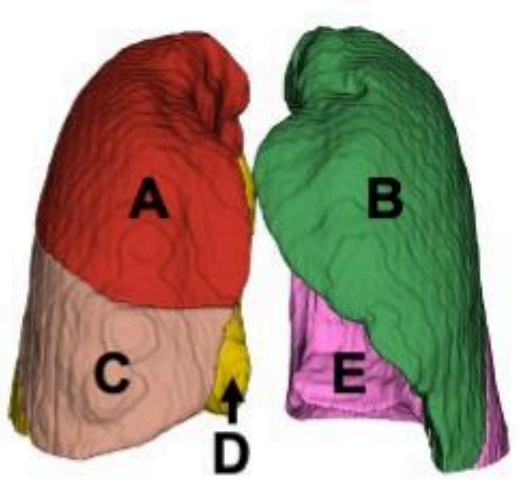

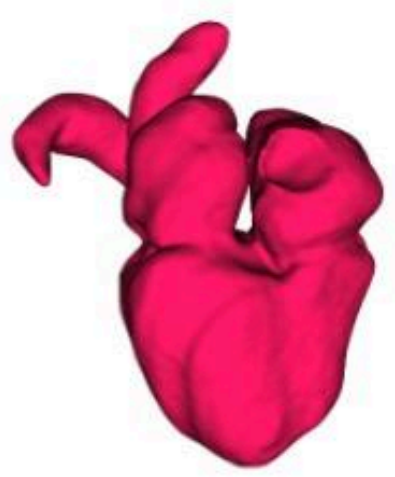

***Figure 2:*** *Anatomical structures included in the final analysis from left to right: thoracic vertebrae T2-T10 (from top to bottom), ribs 3-6, sternum, both lungs, and the heart. The six models segmented each rib individually. During harmonization, we assigned the same color to paired structures, including the ribs. The lungs were segmented into lobes, with the right lung consisting of three lobes (A: right upper lobe, C: right middle lobe, D: right lower lobe) and the left lung consisting of two lobes (B: left upper lobe, E: left lower lobe).*

### *3.3. 3D Slicer extension for comparing segmentation results across models*

We developed *CrossSegmentationExplorer*, a 3D Slicer extension designed to enable efficient comparison of multiple segmentations generated for the same image volume. It is based on the existing 3D Slicer *SegmentationVerification* extension[60]. The extension operates on DICOM images accompanied by segmentations stored as DICOM Segmentation objects. The extension has two main functions: side-by-side visualization of segmentation results and segment-wise inspection of individual segments across all displayed segmentations. The extension is available at: https://github.com/ImagingDataCommons/CrossSegmentationExplorer.

*CrossSegmentationExplorer* allows users to display any selected segmentation in synchronized 2D and 3D views with a single click. When multiple segmentations are selected, the corresponding sets of views are arranged either vertically or horizontally, ensuring that equivalent views are always aligned for direct comparison. This avoids the otherwise time-consuming process of manually creating and linking multiple views in 3D Slicer.

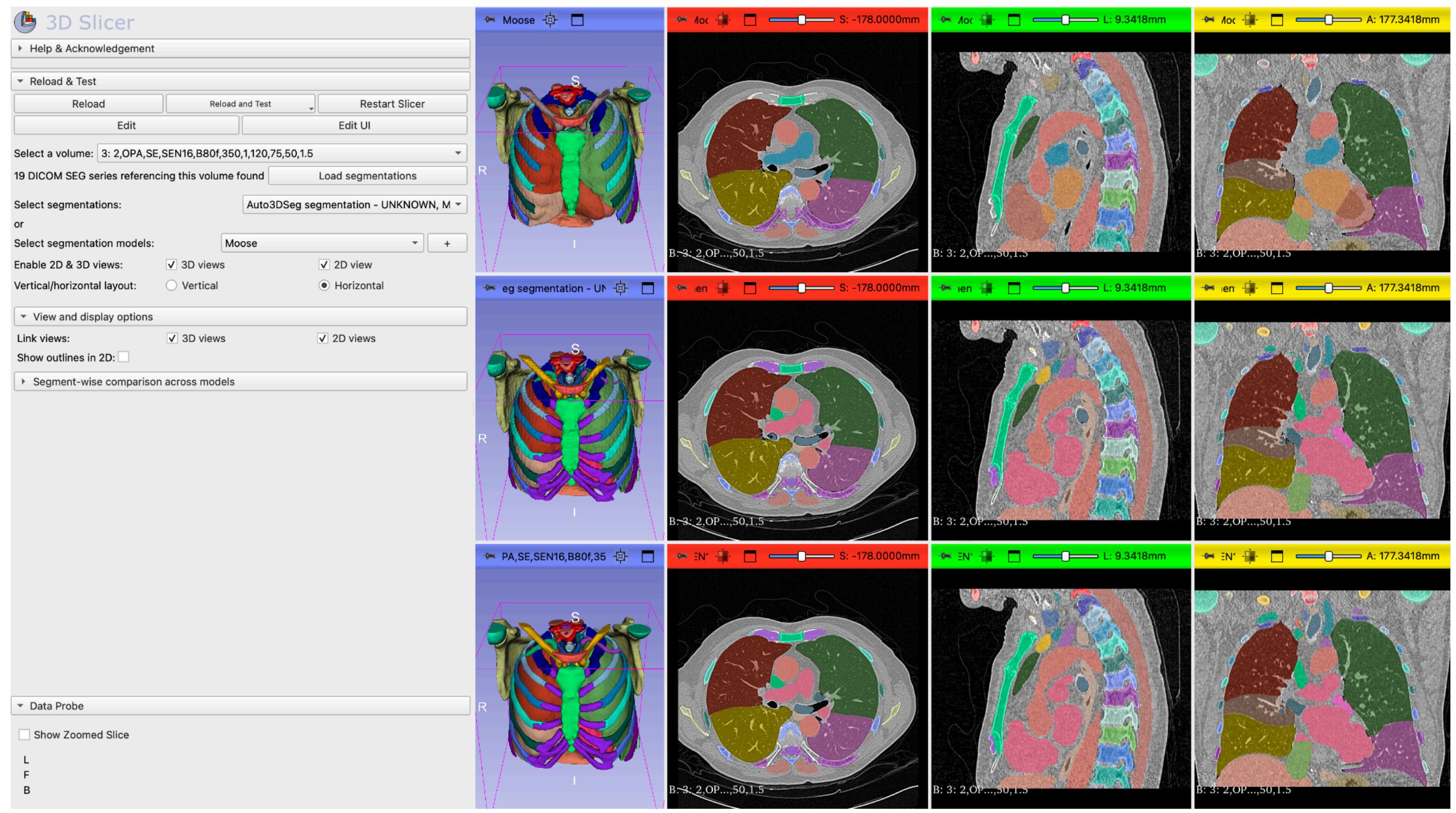

***Figure 3:*** *Example of segmentation visualization in the CrossSegmentationExplorer. After selecting the image volume, segmentation models or segmentation files are chosen from the drop-down menus on the left. For each selected model and file, the corresponding segmentations are automatically displayed in one 3D view and three orthogonal 2D views (axial, sagittal, coronal), with one row of views per selected model. In this example, two segmentation files (Auto3DSeg and MultiTalent) and one segmentation model (MOOSE) have been selected. The Moose segmentation model groups all segmentation files whose series description contains the keyword "MOOSE" or "Moose," and displays them together in a single combined view.*

Since no ground truth data is available, differences between the six segmentation models are especially important, as they may indicate segmentation errors or inconsistencies. Identifying these potential problems relies entirely on visual comparison. *CrossSegmentationExplorer* supports this by making it easy to focus on individual anatomical structures. A dedicated table lists all structures present in the loaded segmentations, and users can select any structure from this table to display it across all models. The extension then shows the chosen structure simultaneously in each segmentation, allowing direct visual comparison of its shape, position, and extent. Navigation controls further simplify the process by enabling users to step through structures one by one, making systematic structure-by-structure comparison straightforward, as illustrated in Figure 3.

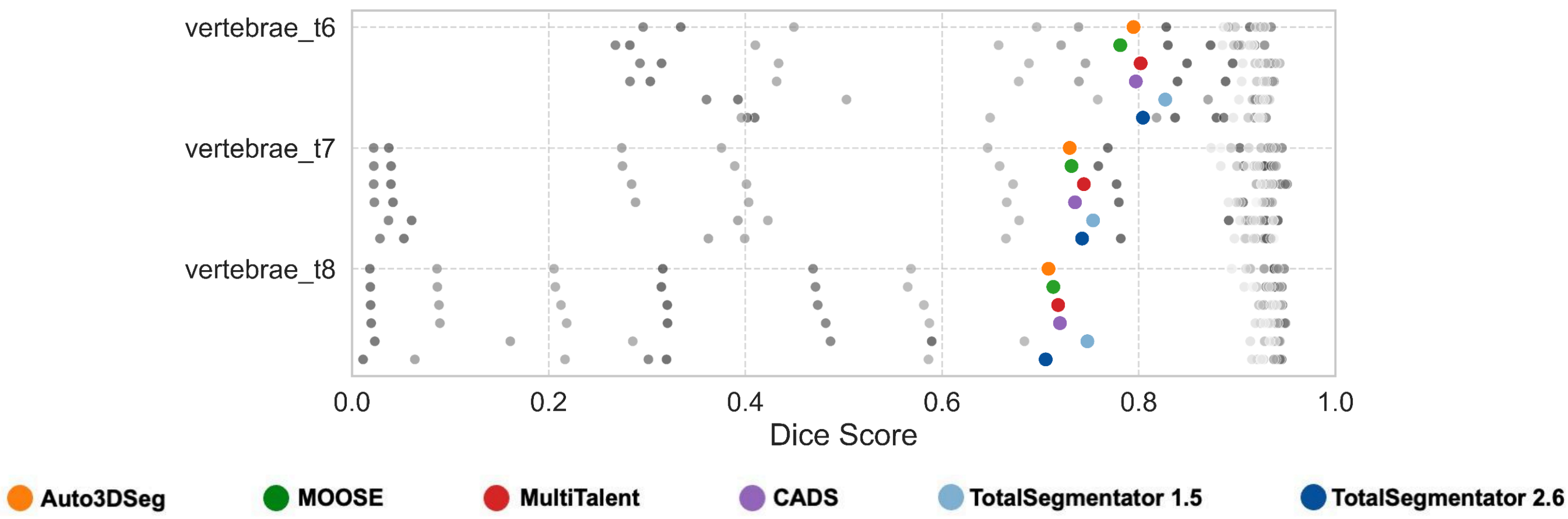

***Figure 4:*** *The plot shows the DSC for three vertebrae as an example of the analysis performed for all anatomical structures and segmentation models in the study. The x-axis shows the DSC value and the y-axis lists the individual structures. Colored points indicate the mean DSC for each model-structure combination, and grey points in the background represent individual DSC measurements for each case.*

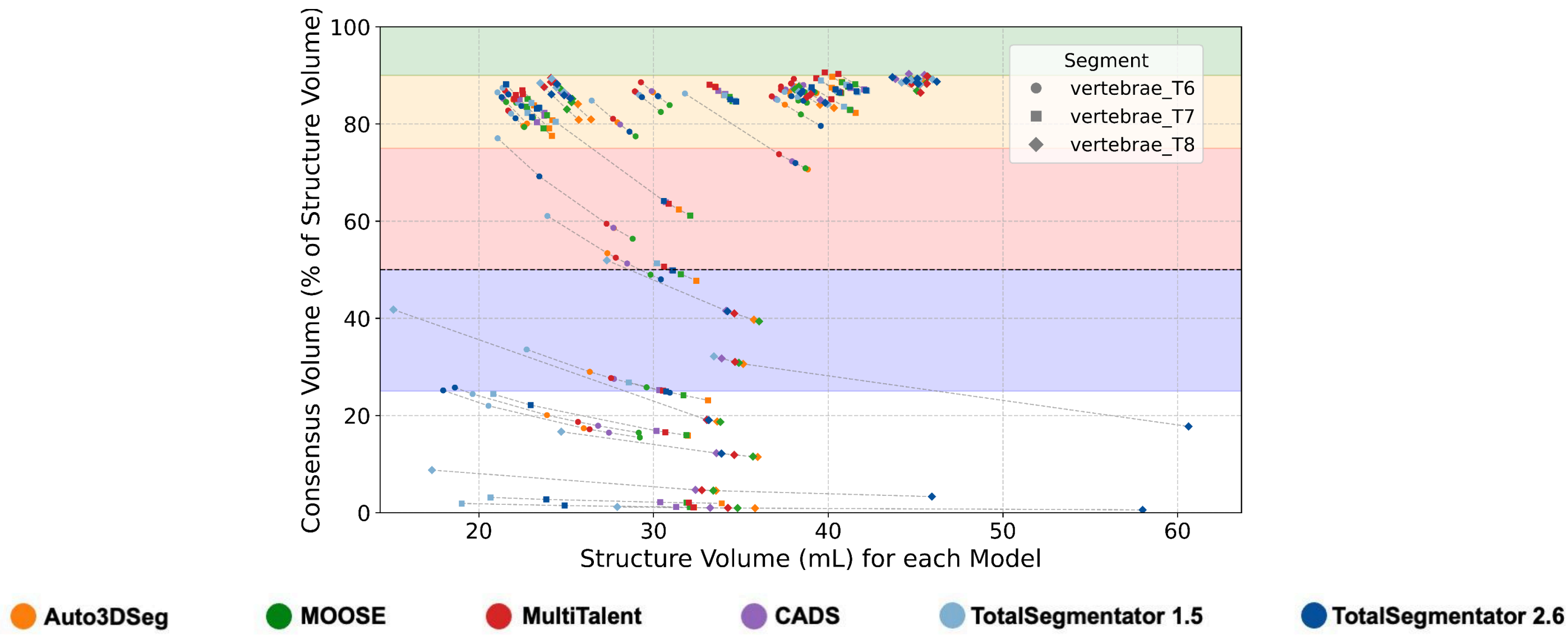

***Figure 5*** *Comparison of the consensus volume with the segmented volume for each structure and model. The x-axis represents the segmented volume, while the y-axis indicates the percentage of the consensus volume relative to the volume of the structure segmented by a specific model. Each point corresponds to a specific segmented structure in one CT scan of the evaluation dataset, produced by a particular segmentation model. Point colors indicate the segmentation model (e.g., green for MOOSE), and point shapes denote the corresponding anatomical structure. Color bands indicate agreement ranges: green for 100–90%, yellow for 90–75%, red for 75–50%, and blue for 50–25%. Points corresponding to the same structure in a given CT scan are connected by dashed lines.*

### *3.4. Visualization of Model Agreement*

Evaluating the 24 anatomical structures in 18 CT scans from our dataset, each segmented by six models, with two metrics produced 2,592 values per metric and over 5,000 individual measurements in total. These values were then visualized in the customized plots we developed for each metric. The interactive versions of all of the plots are available at [https://imagingdatacommons.github.io/segmentation-comparison/](https://imagingdatacommons.github.io/segmentation-comparison/).

To introduce the two plots, we start with the DSC scatterplot, shown in Figure 4, which allows for an assessment of both variability within a structure and the relative performance of the models. Lower DSC values indicate greater deviation from the consensus, making potential outliers easy to identify. For the visualization of the volumes, we also used a scatterplot-based design, as shown in Figure 5. This plot uses color bands to illustrate the level of agreement. Points that fall outside these bands, and therefore have no color background, can be automatically identified as outliers. During the qualitative analysis, these points can be examined first, followed by points within the colored bands, moving from the lowest to the highest agreement range. In addition, dashed lines connecting points for the same structure in a given CT scan allow identification of whether other models achieved better agreement for that structure.

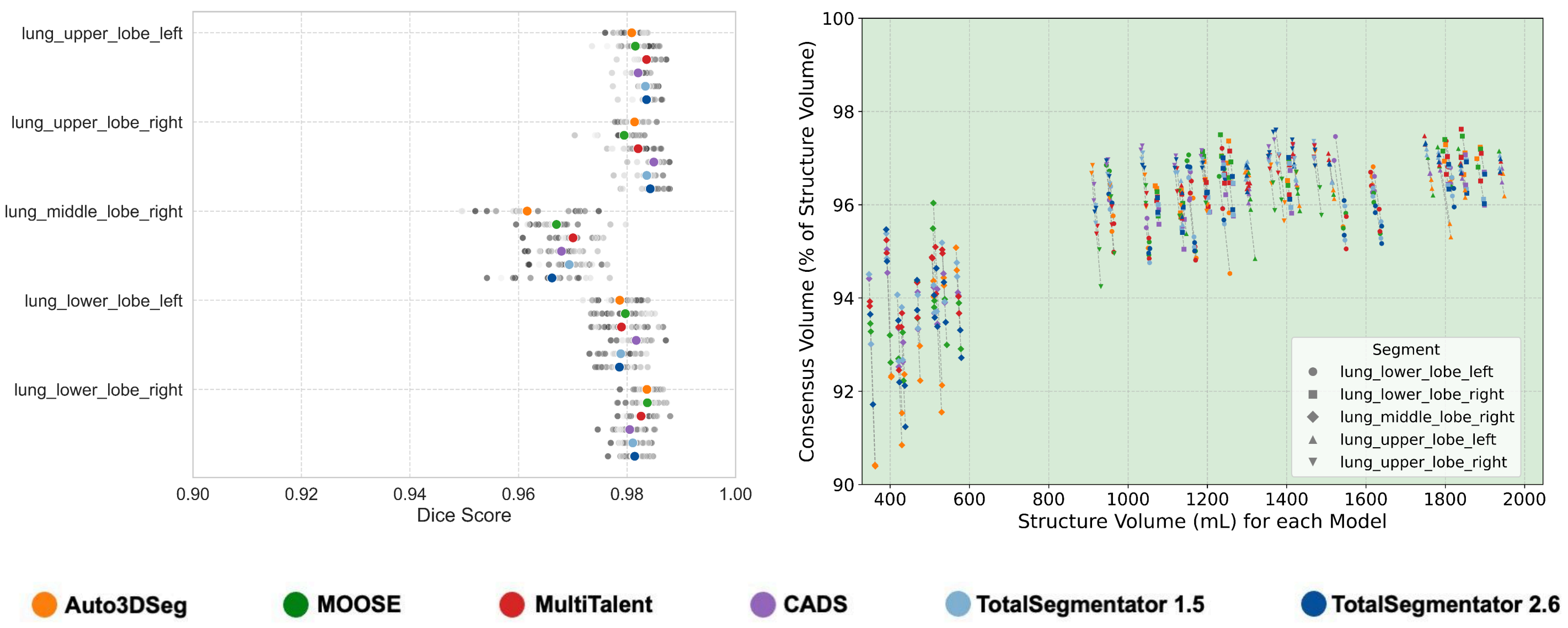

***Figure 6*** *Lung segmentation agreement across models. Both plots show generally high agreement, with DSC above 95% for all lobes (left) and volume overlap above 90% (right). The right middle lobe (the smallest by volume) performs consistently worse than the other lobes in both DSC and volume agreement.*

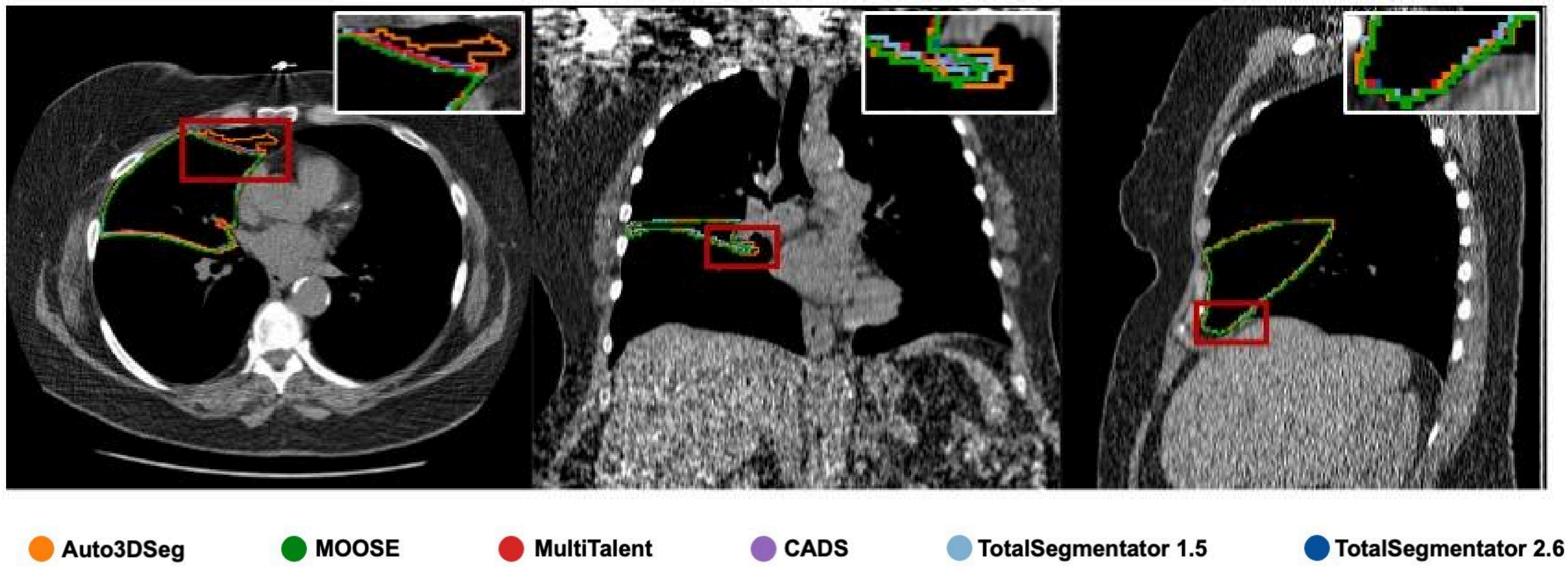

***Figure 7*** *Example 2D visualization of the right middle lobe segmentation for Patient 218890 from our evaluation dataset, shown in the axial, coronal, and sagittal planes.*

### *3.5. Quantitative and Qualitative evaluation*

Our quantitative analysis of the 24 selected anatomical structures produced a large number of consensus DSC and volume agreement values, which were summarized using interactive scatterplots. To simplify interpretation, we organized the structures into five anatomical groups: Lungs (all lobes), Heart (whole heart), Ribs (ribs 3–6), Vertebrae (T2–T10), and Sternum.

The lung segmentations showed best agreement across all models, while the vertebrae showed the lowest agreement. Since all structures, except for the lung lobes, had DSC and volume agreement values below 90%, we performed additional visual comparisons for these structures. For the lungs, we also performed visual spot checks of the segmentations to verify the strong agreement indicated by the metrics.

The following sections provide a detailed analysis of each of these anatomical groups, combining quantitative findings with insights from visual comparisons.

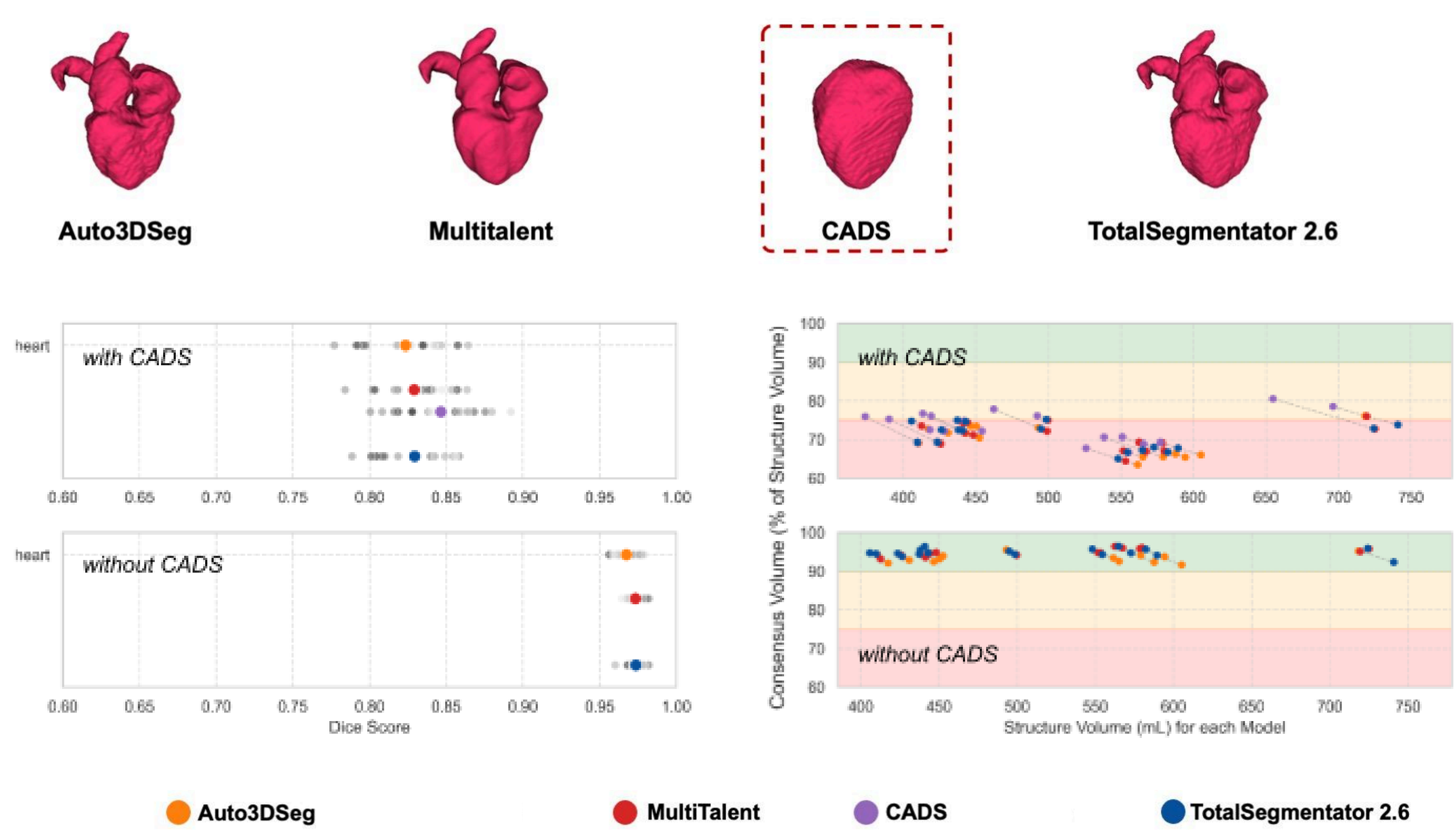

***Figure 8: Top:*** *Visualization of the whole-heart segmentations for Auto3DSeg, MultiTalent, CADS, and TotalSegmentator 2.6. Auto3DSeg, MultiTalent, and TotalSegmentator 2.6 segment the atria and ventricles in more detail and include adjacent blood vessels, whereas CADS represents the heart as a single compact structure that only broadly encompasses the chambers.* ***Bottom:*** *Heart segmentation agreement across the evaluated models. Top row: summary for all of the models demonstrates moderate to good agreement, with DSCs ranging from 75% to 90% (left) and overlap ratios ranging from 60% to 85% (right). Bottom row: summary of the improved agreement after excluding CADS, which uses different conventions for segmenting the heart: the remaining models demonstrate good agreement, with DSCs above 95% (left) and volume overlap exceeding 90% (right).*

### 3.5.1. Lung

The plots in Figure 6 show good agreement in both volume and DSC comparisons across all models and all lung lobes. Across all lung structures and models, DSCs remain consistently above 95%, and volume overlaps exceed 90% in all cases. Visual inspection of the segmentations supports these findings. As shown in Figure 7, even for the middle lobe, the lobe with the lowest agreement among models, only minor boundary differences are present. Because of the good agreement observed across both quantitative metrics and visual comparison, we did not perform further analysis of the lung segmentations.

### 3.5.2. Heart

All six models provide heart segmentations, but they differ in how the heart is represented. *TotalSegmentator* 2.6, *Auto3DSeg*, and *MultiTalent* segment the heart as a single structure, whereas *TotalSegmentator* 1.5 and *MOOSE* segment individual cardiac substructures, such as the ventricles. *CADS* provides both the segmentation of the entire heart as well as separate substructure labels.

The four models that segment the heart as a single structure show only moderate agreement. As shown in the upper row in the two plots of Figure 8, DSCs range from 75% to 90%, and the consensus volume is only 60% to 85% of the individual model volumes.A visual comparison of 3D Segmentations in Figure 8 highlights the cause of the moderate agreement: *CADS* uses a different anatomical definition of the heart than the other models, leading to noticeable differences in the shape of the segmentation. Its segmentation represents a compact structure, likely limited to the ventricles and excluding major vessels. In contrast, the other models include additional anatomical details such as the atria and large vessels (e.g., aorta and pulmonary arteries). With *CADS* excluded from the analysis, the agreement among the remaining models improves significantly, as evident from the bottom row in the two plots of Figure 8.

### 3.5.3. Ribs

All six models segment all twelve rib pairs, with each left and right rib segmented separately. However, rib pairs one and two, as well as seven through twelve, were only partially covered in some CT scans and were therefore excluded from our analysis. As a result, our evaluation includes eight rib structures, rib pairs 3 to 6.

The agreement between models for these rib segmentations ranges from moderate to poor. As shown in the top row in the two plots of Figure 9, DSC varies between 65% and 95%, while the ratio of consensus volume relative to each model's segmentation ranges from 45% to 90%.

The main reason for the lower agreement is segmentation errors in four of the six models, which becomes apparent after visual examination of the results. *TotalSegmentator* 1.5, *TotalSegmentator* 2.6, *Auto3DSeg*, and *MultiTalent* sometimes include parts of neighboring ribs in their segmentations. For example, in the 3D segmentations of Figure 9, the segmentation of the left sixth rib also includes part of the seventh. These errors could be caused by the deficiencies in the training data labels, as noted by Xu et al.[23], while all four models are based on the *TotalSegmentator* dataset.

After excluding the four models affected by these errors, we observed improved agreement among the remaining two models. Nevertheless, agreement remained in the moderate range. As shown in the bottom row in the two plots of Figure 9, the DSC are above 80%, and the consensus covers between 65% and 95% of the volume that is segmented by the individual models, if we consider only *MOOSE* and *CADS*.The remaining disagreement between *MOOSE* and *CADS* results from differences in segmentation coverage, which is likely caused by the differences in the training data (due to the varying conventions used to perform the segmentation, or individual preferences of the experts performing the segmentation). *MOOSE* is the only model that includes that portion of the rib within the costovertebral joints (where the ribs articulate with the vertebrae). When comparing this region across models (see Figure 10), it becomes clear that the four error-prone models fail to include the medial rib portions near the spine, resulting in visible gaps. *CADS* includes a slightly larger portion of the rib within the costovertebral joints, but does not reach the coverage of the *MOOSE* segmentation. As a result, *MOOSE* consistently produces higher rib volumes than *CADS*.

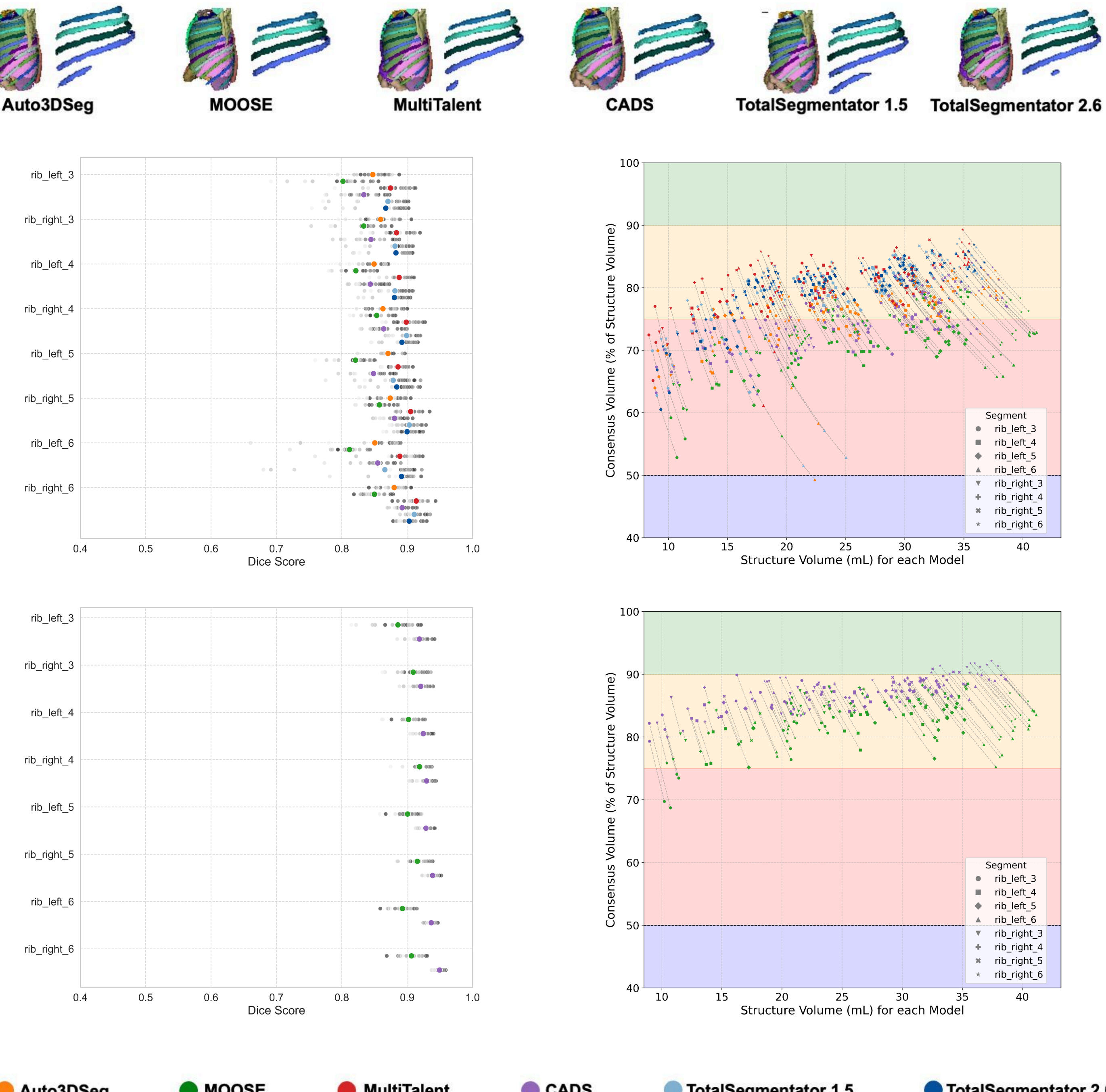

***Figure 9: Top:*** *Example of segmentation errors in rib structures for Patient 218890 from the evaluation dataset. Although all four models with rib segmentation errors (TotalSegmentator 1.5, TotalSegmentator 2.6, Auto3DSeg, and MultiTalent) were trained on the same dataset, they produce different segmentation errors. For each model, the left image shows all segmented structures, including those excluded from the analysis, to illustrate the extent of incorrectly segmented ribs. The right image for each model displays only the ribs included in our analysis.* ***Bottom:*** *Rib segmentation agreement across models. Top row: summary for all models demonstrates moderate to poor agreement, with DSCs ranging from 65% to 95% (left) and volume agreement between 45% and 90% (right). Rib 6 left performs worst as measured both using DSCs and volume agreement, followed by rib 3 pair. Bottom row: summary after excluding TotalSegmentator 1.5, TotalSegmentator 2.6, Auto3DSeg, and MultiTalent due to segmentation errors. The comparison now includes only MOOSE and CADS, which did not rely on the TotalSegmentator training dataset. Agreement improves notably, with DSCs above 80% and volume agreement ranging from 65% to 95%.*

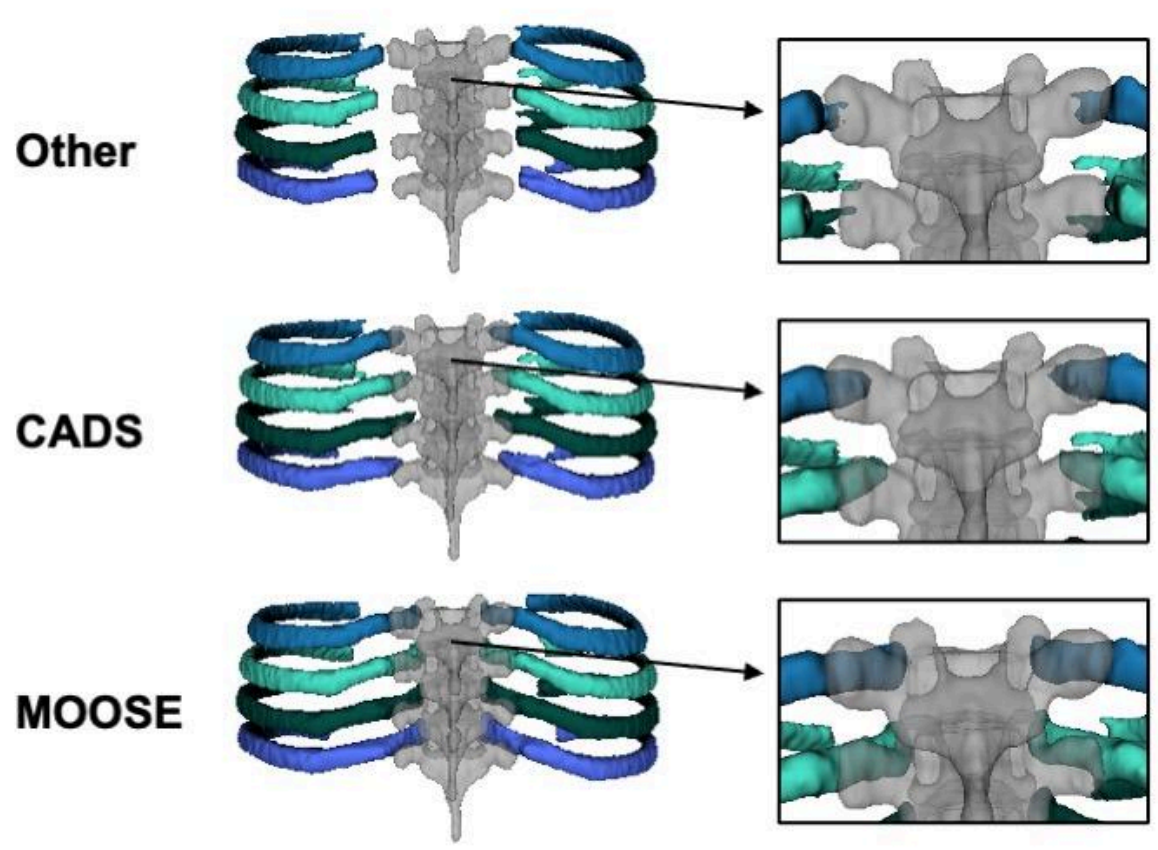

***Figure 10:*** *Illustration of the differences in the gap observed between the rib and vertebra segmentation across models. Left: examples from TotalSegmentator 1.5 (representative of the other four models), CADS, and MOOSE. All models except for MOOSE and CADS show large gaps between the ribs and the vertebrae, CADS reduces this gap, and MOOSE fully includes the rib within the costovertebral joints. Right: an enlarged view of the space between ribs 3 and 4 and vertebrae T3 and T4, highlighting improved segmentation coverage by MOOSE segments the entire joint, unlike CADS.*

### 3.5.4. T2-T10 vertebrae

As with the lungs and ribs, all six models segment the thoracic vertebrae. Our analysis focuses on vertebrae T2 through T10, which form the thoracic cage in combination with the ribs and sternum. We excluded the remaining vertebrae because they were not fully included in some CT scan series.

The top row in the two plots of Figure 11 shows poor agreement between the models in vertebral segmentations, especially for T6 through T9. There is a wide variation in DSCs and the proportion of consensus volume to model volume, ranging from 0% to 95%.

This low agreement is primarily due to segmentation errors in the same four segmentation models that also had difficulties segmenting the ribs: *TotalSegmentator* 1.5, *TotalSegmentator* 2.6, *MultiTalent* and *Auto3DSeg*. As for the rib segmentations, these errors are likely caused by issues in the *TotalSegmentator* training dataset, which is known to contain problematic vertebrae segmentations (https://github.com/wasserth/*TotalSegmentator*/issues/394).

We observed three main types of segmentation errors that are shown in the 3D Segmentations of Figure 11:

1. Inclusion of parts of adjacent vertebrae.
2. Merging of multiple vertebrae under a single label.
3. Incorrect assignment of the vertebra label.

In contrast, *CADS* and *MOOSE* did not show any drastic errors in vertebral segmentation on our evaluation dataset.

When limiting the comparison to *CADS* and *MOOSE*, agreement improves, with *CADS* and *MOOSE* showing good agreement for vertebrae. As shown in the bottom row in the two plots of Figure 11, DSCs exceed 90% for all vertebrae, and consensus volumes are above 85%.

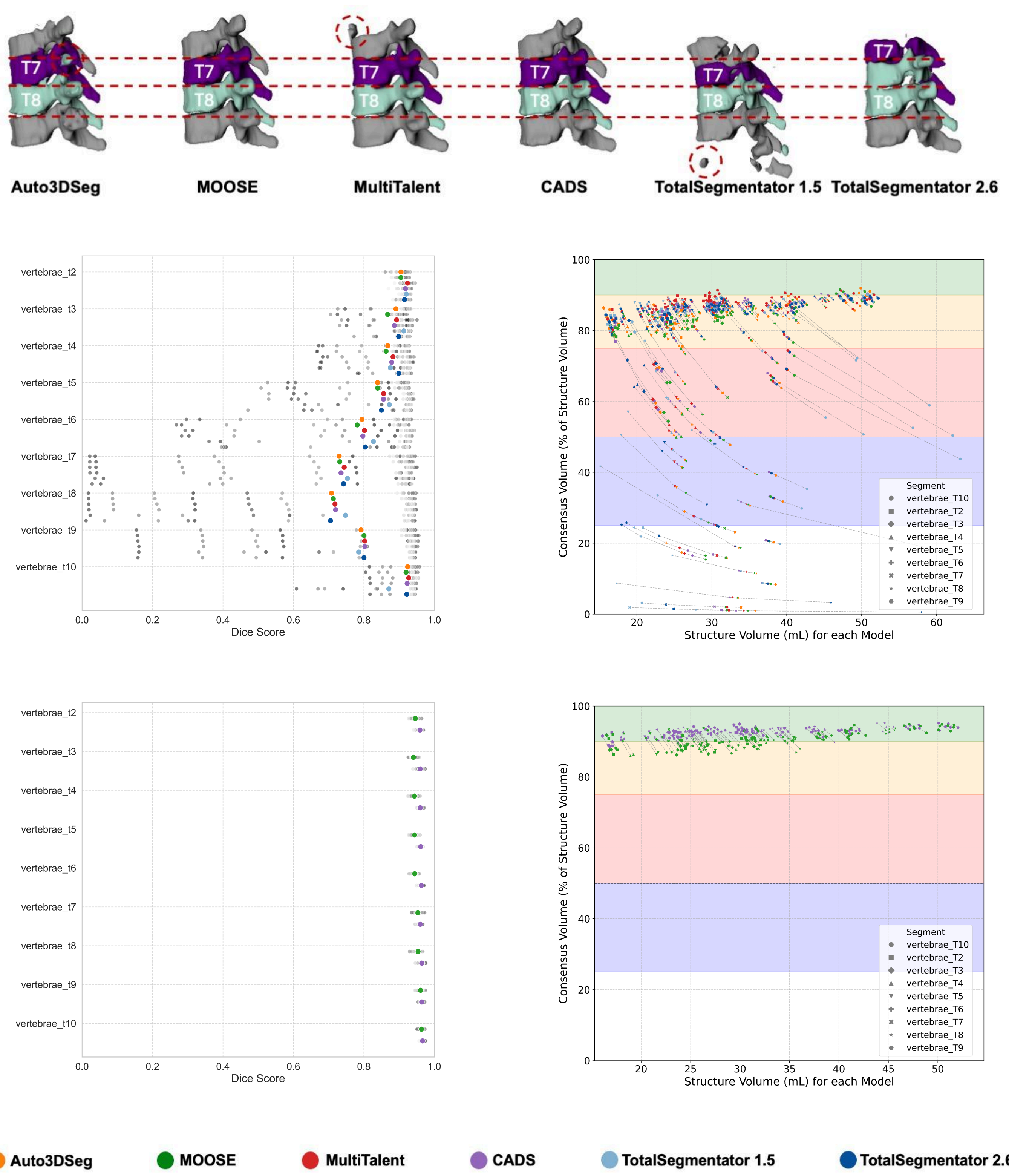

***Figure 11: Top:*** *Examples of segmentation errors for vertebra T8 (light green) in Patient 100004 from the evaluation dataset. No segmentation errors were observed for MOOSE and CADS. For Auto3DSeg, MultiTalent, and TotalSegmentator 1.5, inclusion of parts of adjacent vertebrae can be seen (highlighted with red circles). In TotalSegmentator 2.6, fusion of multiple vertebrae under a single label is visible for T8, and incorrect label assignment occurs for T7 (purple), as a result of the fusion error.* ***Bottom:*** *Vertebral segmentation agreement across models. Top row: Agreement across all analyzed models shows large*

*variability, with DSCs ranging from 0% to 95% (left) and volume agreement between 0% and 95% (right). Vertebrae T7 and T8 show the lowest agreement, with some DSC and volume overlap values close to zero. Vertebrae T5 to T9 also show poor performance overall, while T2 to T4 and T10 perform better, with DSCs above 60% and volume agreement exceeding 40%. Bottom row: Vertebra segmentation agreement across the MOOSE and CADS models, after excluding TotalSegmentator 1.5, TotalSegmentator 2.6, Auto3DSeg, and MultiTalent due to segmentation errors. Agreement improves significantly compared to the top row, with DSCs exceeding 90% for all vertebrae and volume agreement above 85%.*

The volume agreement is slightly lower than the DSCs, which can be explained by the fact that *MOOSE* segments larger vertebral volumes than *CADS*. This difference is especially noticeable in the intervertebral spaces (which should be excluded from the volume). *MOOSE* includes less spacing between adjacent vertebrae (intrudes more into the intervertebral space), leading to larger segmented volumes overall. However, based on the expert review (R.K.), the CT resolution between slices in our evaluation dataset is too thick to clearly determine which segmentation is anatomically more accurate, so the ground truth is not known.

### 3.5.5. Sternum

Five of the six models in our analysis segment the sternum: *TotalSegmentator* 2.6, *Auto3DSeg*, *MOOSE*, *MultiTalent*, and *CADS*. *TotalSegmentator* 1.5 is the only model that does not segment this structure.

Overall agreement between the five models is moderate, as shown in the top row in the two plots of Figure 12. DSCs exceed 85%, and the consensus volume accounts for more than 75% of the individual segmentations. Excluding the *CADS* results leads to a slight improvement, though this is not significant (see the second row in the two plots of Figure 12).

The 3D Segmentations of Figure 12 show that there are systematic differences in how the models capture the transition from the sternal body to the xiphoid process in the segmentations of the sternum. *TotalSegmentator* 2.6 and *Auto3DSeg* segment the xiphoid process as a longer, straighter structure before ending in a slightly rounded tip. In contrast, *MOOSE* and *MultiTalent* depict the xiphoid process as shorter and more rounded. The models within each pair show good agreement, as shown in the third and bottom row in the two plots of Figure 12.

*CADS*, in contrast, segments the xiphoid process differently from all other models and assigns less volume to the sternum body, which may explain its lower overall agreement with other models.4. Discussion

We presented initial results of a systematic comparison of AI models for automatic segmentation of chest organs, in absence of ground truth annotations. In the process of performing this comparison we established procedures and tools that rely on established standards and open source software to facilitate similar comparisons. On the example of 18 CT scans from the NLST collection segmented by six models across 24 structures, resulting in 2,592 individual segmentations in our evaluation dataset, we demonstrated how harmonized segmentation results can be efficiently navigated allowing to evaluate consistency of segmentations across models and identify the most relevant cases for comparison and visual inspection. In turn, we demonstrated how uniform, standard-compliant encoding of the segmentation results makes it easy to visualize those in the zero-footprint OHIF Viewer and compare segmentation of the same structure across models using the developed open-source 3D Slicer CrossSegmentationExplorer module. Using this approach, we identified segmentation deficiencies in four of the six evaluated models. We publicly share the data harmonization tools, resulting DICOM datasets, interactive plots and 3D Slicer extension to ensure reproducibility, stimulate adoption and further development of this approach.

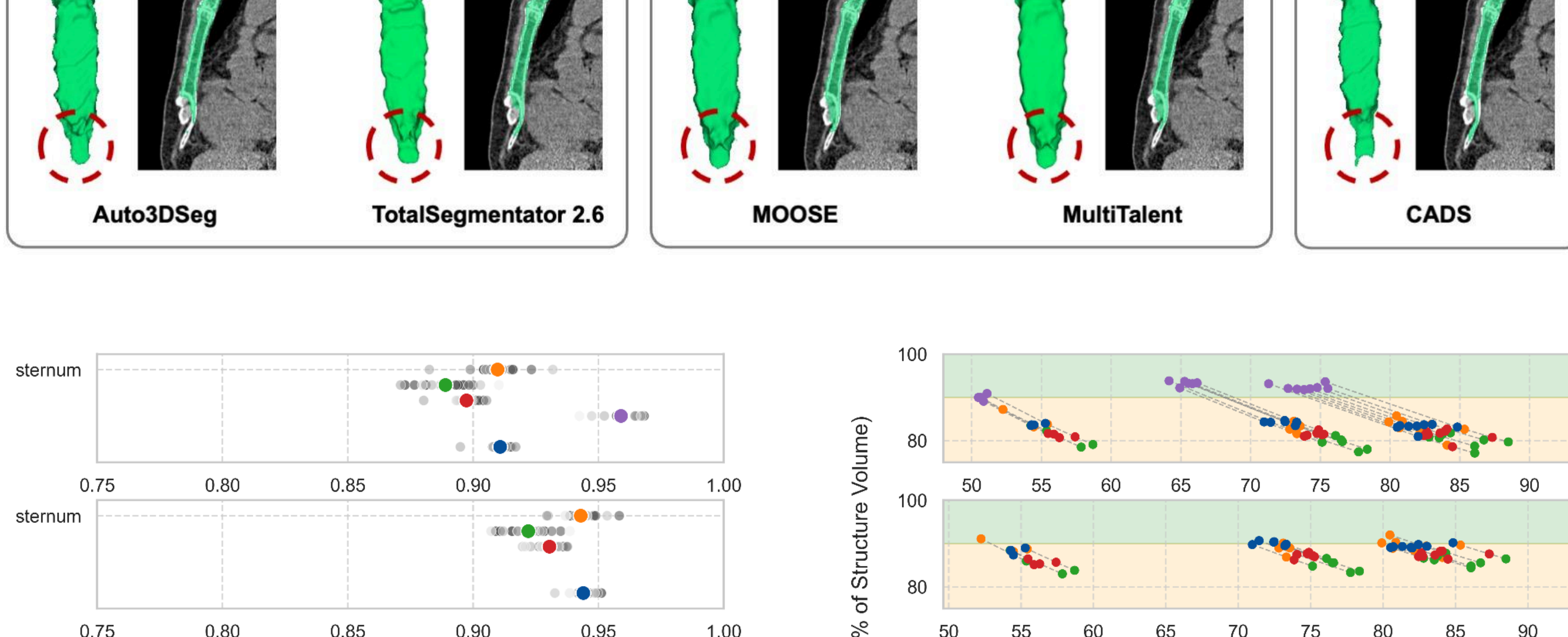

***Figure 12: Top:*** *Comparison of sternum segmentations from all models for PatientID 218808 in the evaluation dataset. For each model, the figure shows a 3D rendering of the segmented sternum alongside the corresponding 2D segmentation in a sagittal view. The models within each outlined pair show similar segmentation patterns. Auto3DSeg and TotalSegmentator 2.6 represent the xiphoid process (highlighted with red circles) as longer and straighter, ending in a slightly rounded tip, while MOOSE and MultiTalent depict it as shorter and more rounded. CADS segments the xiphoid process differently from all other models and assigns less volume to the sternal body.* ***Bottom:*** *Sternum segmentation agreement across models using DSC (left column) and volumetric overlap (right column). Top row: Comparison across all models, showing moderate overall agreement with DSCs above 85% for all cases and volume agreement between 75% and 95%. Second row: Excluding CADS slightly improves agreement, with DSCs now exceeding 90% and volume agreement generally above 80%. Third row: When CADS is excluded and models are compared pairwise, Auto3Dseg and TotalSegmentator 2.6 demonstrate good agreement, with DSCs above 95% and volume agreement above 90%. Bottom row: MOOSE and MultiTalent also show good agreement, with DSCs exceeding 95% and volume agreement above 90%.*

Our framework aims to assist in informed decision making on model selection with or without ground truth segmentations. Importantly, it is universally applicable to any other dataset and any chosen segmentation model. While this development was motivated by the need to compare publicly available segmentation tools for the purposes of annotating the NLST collection, the same approach can also be applied to other datasets and use-cases. We hope that improved understanding of strengths and weaknesses of publicly available segmentation models will stimulate their adoption and secondary use of the segmentation results to help provide valuable information about the study population, including organ and structure volumes, anatomical variability, and disease-related alterations.

Both our systematic comparison and the "toolkit" itself have some limitations. Our analysis was restricted to 18 CT scans out of the more than 126,000 scans available in NLST, to explore the utility of the tools, and demonstrate the proof of concept. The selected sample does not reflect the demographic diversity of the original dataset (e.g, our subset included only white participants), or the heterogeneity of the image acquisition equipment or image reconstruction approaches. We do not know if the differences we observed across the models will generalize, and it is our plan to extend the evaluation to a representative sample of the NLST collection. The NLST dataset is limited to screening chest CT scans. As a result, important abdominal organs are excluded from the analysis. The image quality characteristics of low dose screening CTs may not be generalizable to more routine CT scan protocols. Analyzing these factors would require applying our approach to other datasets. However, since our tools are both dataset- and model-agnostic, such an analysis can now be performed more easily. It remains to be seen how scalable are the types of plots we utilized in our analysis when used with the larger number of samples (we could have tested this using an artificial sample).

Another limitation is that our analysis relies on inter-model consensus, which means that errors or problems shared by all models are not detected. In such cases, the models would still appear to be in perfect agreement, even though all of them are wrong. Such errors could only be detected through ground truth annotations or clinical expert review, which is not practical given the large number of segmentations. For our purpose of identifying the most suitable model for the dataset, this limitation is less critical, as in such cases all available models would fail to correctly segment the structure.

Building on this point, another limitation lies in how we quantified model performance in relation to the consensus. Specifically, we calculated the DSC similarity and volume agreement between each model and the consensus segmentation. While this approach is straightforward and easy to interpret, it has clear restrictions. More advanced methods, such as Simultaneous Truth and Performance Level Estimation (STAPLE)[30], could provide a more robust estimate by explicitly modeling the probability of error for each model and combining their outputs accordingly. However, STAPLE also has important limitations. For example, if the majority of models segment a structure incorrectly, the resulting estimate can still be biased toward the wrong solution. Moreover, STAPLE assumes that models make errors independently, which does not hold in our setting, as four of the evaluated models were at least partly trained on the same datasets and are therefore likely to share systematic errors. For these reasons, we considered STAPLE beyond the scope of this study.

A further limitation relates to our Slicer *CrossSegmentationExplorer* module. At present, the module was tested only with DICOM CT images and DICOM SEG files, but should work for any segmented cross-sectional radiology modality (CT, MR or PET) as long as the data is represented in standard-compliant DICOM representation. Therefore, it is not possible to use the NIfTI outputs of the models directly, since the metadata required for selecting segmentations of the matching structures is unavailable in a standard format and therefore is not readily ingested. To add a new segmentation model, a structure harmonization mapping must first be created. While this is straightforward with our approach, and facilitated by using a standard DICOM format, it nonetheless requires additional effort before the data can be used by the module. It would be highly desirable that future segmentation modules output standard DICOM SEG files with the structure metadata populated with standard coded values for the segment labels, thus improving interoperability and simplifying comparisons with alternative tools.

To continue the analysis and establish the generalizability of the early findings, we plan to extend the study to a larger and more representative subset of the NLST dataset. This will allow us to assess whether the errors observed in the initial sample persist, whether additional segmentation issues can be identified in a broader evaluation, and whether the repeat studies of the same patient can be leveraged to better assess reproducibility. To further strengthen the analysis, we also plan to incorporate STAPLE as an additional method for consensus estimation. Based on the proposed extended analysis, we aim to identify the most suitable model for NLST and use it to generate segmentations for the entire dataset, which we will then publish on IDC. With these segmentations, we want to explore what new insights about the underlying populations of the NLST cohort can be obtained from the derived segmentation data. The same principles could also be applied to other IDC datasets, including other modalities (numerous models are now available for organ segmentation from Magnetic Resonance Imaging (MRI)[61,62]), enabling broader availability of large-scale segmentations.

# 5. Conclusion

We have presented a toolkit for systematic comparison of segmentation models that addresses the challenge of evaluating large datasets without ground truth annotations or time-consuming expert reviews. Starting from the raw outputs of the segmentation models, we first harmonized all labels and then encoded the results into standard DICOM SEG format. We then applied two quantitative metrics and visualized the results in interactive plots that made it possible to quickly identify outliers and preselect cases for further analysis. These cases were then inspected with our new 3D Slicer CrossSegmentationExplorer module, and unusual findings were further reviewed by a clinical expert (R.K.). Using this workflow, we detected segmentation errors in the ribs and vertebral structures in four (*TotalSegmentator* 1.5, *TotalSegmentator* 2.6, *Auto3DSeg*, *MultiTalent*) of the six evaluated models, the four known to share the same training data.

With this framework, we showed that at least some of the segmentation errors can be detected even in models that otherwise generally perform well, even in absence of ground truth annotations. While our approach cannot capture every possible segmentation error, and depends on inter-model agreement, it is easy to apply, scalable to large datasets, and effective in highlighting problematic cases. We hope that by releasing our tools and evaluation data, we enable the broader community to apply and adapt this workflow to other datasets lacking ground truth annotations and to make an informed model selection. Further, these results can help developers of the AI models explore existing limitations and develop better segmentation tools and better training datasets.

# Disclosures

The authors have no conflicts of interest to declare.

# Code, Data and Materials availability statement

Scripts involved in the processing of the data are available at https://github.com/ImagingDataCommons/segmentation-comparison. *CrossSegmentationVerification* 3D Slicer extension developed is available at https://github.com/ImagingDataCommons/CrossSegmentationExplorer. The interactive plots summarizing results are available at https://imagingdatacommons.github.io/segmentation-comparison/.

This study involved analysis of the data publicly available as part of the NLST[15] and TotalSegmentator-CT-Segmentations[39] collections, downloaded from Imaging Data Commons. The analysis results produced in the course of the analysis are shared publicly[59].

# Acknowledgments

The study described in this paper started as a project at the NA-MIC Project Week[63] 42, see https://projectweek.na-mic.org/PW42_2025_GranCanaria/Projects/ReviewOfSegmentationResultsQualityAcrossVariousMultiOrganSegmentationModels/. This work was supported by the National Cancer Institute, National Institutes of Health, under Task Order No. HHSN26110071 under Contract No. HHSN261201500003l.

# Appendix

We provide a summary of the key acquisition-related and demographic metadata for the CT images analyzed in the study in Table 1.

| PatientID | SeriesInstanceUID | Manufacturer | Manufacturer Model | Convolution Kernel | Slice Thickness | Pixel Spacing | Age | Sex |
|---|---|---|---|---|---|---|---|---|
| 100002 | 1.2.840.113654.2.55.257926562693607663865369179341285235858 | GE | LightSpeed Plus | STANDARD | 2.5 | 0.703 | 66 | M |
| 100002 | 1.2.840.113654.2.55.229650531101716203536241646069123704792 | GE | LightSpeed Plus | LUNG | 2.5 | 0.703 | 66 | M |
| 100002 | 1.2.840.113654.2.55.283399418711252976131557177419186072875 | GE | LightSpeed Plus | STANDARD | 2.5 | 0.703 | 66 | M |
| 100002 | 1.2.840.113654.2.55.21461438679308812574178613217680405233 | GE | LightSpeed Plus | LUNG | 2.5 | 0.703 | 66 | M |
| 100002 | 1.2.840.113654.2.55.122344168497038128022524906545138736420 | GE | LightSpeed Plus | LUNG | 2.5 | 0.703 | 66 | M |
| 100002 | 1.2.840.113654.2.55.97114726565566537928831413367474015470 | GE | LightSpeed Plus | STANDARD | 2.5 | 0.703 | 66 | M |
| 100004 | 1.2.840.113654.2.55.195946682403058845904768502826466194287 | GE | LightSpeed Plus | LUNG | 2.5 | 0.801 | 60 | M |
| 100004 | 1.2.840.113654.2.55.221581533879834196356530174246594024639 | GE | LightSpeed Plus | STANDARD | 2.5 | 0.801 | 60 | M |
| 100004 | 1.2.840.113654.2.55.71263399928421039572326605504649736531 | GE | LightSpeed Plus | LUNG | 2.5 | 0.801 | 60 | M |
| 100004 | 1.2.840.113654.2.55.79318439085250760439172236218713769408 | GE | LightSpeed Plus | STANDARD | 2.5 | 0.801 | 60 | M |
| 100004 | 1.2.840.113654.2.55.304075689731327662774315497031574106725 | TOSHIBA | Aquilion | FC30 | 2.0 | 0.801 | 60 | M |
| 100004 | 1.2.840.113654.2.55.191661316001774647835097522264785668378 | TOSHIBA | Aquilion | FC01 | 2.0 | 0.801 | 60 | M |
| 218808 | 1.3.6.1.4.1.14519.5.2.1.7009.9004.315696884435641630605419115484 | SIEMENS | Sensation 16 | B30f | 1.0 | 0.684 | 74 | M |
| 218808 | 1.3.6.1.4.1.14519.5.2.1.7009.9004.135383252566920035150987356231 | SIEMENS | Sensation 16 | B80f | 1.0 | 0.684 | 74 | M |
| 218890 | 1.3.6.1.4.1.14519.5.2.1.7009.9004.230644512623268816899910856967 | GE | LightSpeed Ultra | STANDARD | 2.5 | 0.590 | 73 | F |

| 218890 | 1.3.6.1.4.1.14519.5.2.1.7009.9004.330739122093904668699523188451 | GE | LightSpeed Ultra | BONE | 2.5 | 0.590 | 73 | F |
| --- | --- | --- | --- | --- | --- | --- | --- | --- |
| 218890 | 1.3.6.1.4.1.14519.5.2.1.7009.9004.690272753571338193252806012518 | GE | LightSpeed Ultra | BONE | 2.5 | 0.703 | 73 | F |
| 218890 | 1.3.6.1.4.1.14519.5.2.1.7009.9004.310718458447911706151879406927 | GE | LightSpeed Ultra | STANDARD | 2.5 | 0.734 | 73 | F |

**Table 1.** *Acquisition parameters for each CT volume included in the evaluation dataset. Each row corresponds to a single CT series. Row background indicates whether consecutive rows corresponding to CT series belong to the same imaging study. The table includes information on scanner manufacturer and model, reconstruction kernel, slice thickness, in-plane pixel spacing, as well as patient age and sex. The corresponding CT images can be downloaded using idc-index Python package (https://github.com/ImagingDataCommons/idc-index) using command "idc download <SeriesInstanceUID>".*